\documentclass[aps,prb,reprint,showpacs,superscriptaddress]{revtex4-1}

\usepackage[usenames,dvipsnames]{color}
\usepackage{amsmath}    % need for subequations
\usepackage{graphicx}   % need for figures
\usepackage{color}      % use if color is used in text
\usepackage{subfigure}  % use for side-by-side figures
\usepackage{hyperref}   % use for hypertext links, including those to external documents and URLs
\usepackage[utf8]{inputenc}
\usepackage{multirow}
\usepackage{siunitx}

\usepackage[T1]{fontenc}
\usepackage[utf8]{inputenc}
\usepackage[english]{babel}
\usepackage{booktabs}
\usepackage{soul}
\usepackage{epstopdf}

	\usepackage{color}
	\usepackage{MnSymbol}
	\usepackage{dashrule}

	\usepackage{hyperref}
	\hypersetup{
	    colorlinks,
	    citecolor=blue,
	    linkcolor=blue,
	    urlcolor=blue}

\begin{document}
%Antwortfunktion
%EELS: Liang_PhilosophicalMagazine_1969_19_161_1031

\title{Investigation of the dispersion and the effective masses of excitons in bulk $2H$-MoS$_2$ using transition electron energy-loss spectroscopy}
\author{Carsten Habenicht}
\author{Martin Knupfer}
\email{m.knupfer@ifw-dresden.de}
\author{Bernd Büchner}
\affiliation{IFW Dresden, P.O. Box 270116, D-01171 Dresden, Germany}
\date{\today}

\begin{abstract}

We have investigated the electronic excitations in bulk $2H$-MoS$_2$ using electron energy-loss spectroscopy. The electron energy-loss spectra in the $\Gamma$M and $\Gamma$K directions were measured for various momentum transfer values. The results allow the identification of the A$_1$ and B$_1$ exciton peaks and in particular their energy-momentum dispersion. The dispersions exhibit approximately quadratic upward trends and slight anisotropies in the $\Gamma$M and $\Gamma$K directions. The fitted energy-momentum transfer functions allow the estimation of the effective masses of the excitons which are in close proximity to predicted values.
\end{abstract}


\maketitle

\section{INTRODUCTION}
%Crystal structure
MoS$_2$ is a transition-metal dichalcogenide which has received significant scientific interest due to its unique mechanical and electronic properties arising largely from its layered crystalline structure. Besides its traditional use as a lubricant, it has been investigated for applications in photovoltaics, batteries, catalysis, and electronics\cite{Xu_Chemicalreviews_2013_113_5_3766,Ganatra_ACSnano_2014_8_5_4074}.
 Similar to graphene, the crystals are made up of parallel layers loosely linked by Van-der-Waals forces and stacked in the direction of the $c$ axis (for axis orientation and visualization of the crystal structure see Ref.~\citenum{Wilson_AdvancesinPhysics_1969_18_73_193}). Each layer consists of a molybdenum sheet sandwiched between two sulfur sheets. This investigation was performed on the $2H$ polytype (space group: P6$_3$/mmc, D$_{6h}^4$) in which the atoms within the layers have a trigonal prismatic coordination with molybdenum at the center ionically-covalently bound \cite{Heda_JournalofPhysicsandChemistryofSolids_2010_71_3_187} to six sulfur atoms \cite{Dickinson_JournaloftheAmericanChemicalSociety_1923_45_6_1466}. Each unit cell contains two sulfur-molybdenum-sulfur layers that are rotated by 60$^\circ$ with respect to each other and offset in such a way that the molybdenum atoms in one layer are collinear in the $c$ direction with the sulfur atoms in the adjacent layers.

%Electronic properties
The weak intralayer bonding gives rise to quasi two-dimensional electronic properties of the material. Bulk MoS$_2$, which was used in this work, is an semiconductor with an indirect band gap of approximately 1.2 - 1.3 eV.\cite{Mattheiss_PhysicalReviewB_1973_8_8_3719,Kam_TheJournalofPhysicalChemistry_1982_86_4_463,Baglio_JournaloftheElectrochemicalSociety_1982_129_7_1461,Komsa_PhysicalReviewB_2012_86_24_241201, Cheiwchanchamnangij_PhysicalReviewB_2012_85_20_205302} Its valence band maximum is located at the $\Gamma$ point of the Brillouin zone and the conduction band minimum close to the halfway point between $\Gamma$ and K. The bulk material also exhibits a larger direct gap at the K point. \cite{Baglio_JournaloftheElectrochemicalSociety_1982_129_7_1461,Kam_TheJournalofPhysicalChemistry_1982_86_4_463,Peelaers_PRB_2012_86_24_241401,Jiang_TheJournalofPhysicalChemistryC_2012_116_14_7664,Komsa_PhysicalReviewB_2012_86_24_241201}
As the number of crystal layers is reduced down to a single layer, the band gap shifts from an indirect to a direct band gap at K due to the elimination of interlayer interactions and quantum confinement. \cite{Mak_PhysicalReviewLetters_2010_105_13_136805, Kuc_PhysicalReviewB_2011_83_24_245213,Yun_PhysicalReviewB_2012_85_3_33305,Splendiani_Nanoletters_2010_10_4_1271, 
Ellis_AppliedPhysicsLetters_2011_99_26_261908, Mak_PhysicalReviewLetters_2010_105_13_136805, Han_PhysicalReviewB_2011_84_4_45409,Jin_PRL_2013_111_10_106801, Kadantsev_SolidStateCommunications_2012_152_10_909,
Molina-Sanchez_PRB_2013_88_4_45412}

%Excitons
An interesting feature of MoS$_2$ is the existence of a number of excitonic transitions in the visible region. The energetically lowest ones are the A$_n$ and B$_n$ exciton series located near 2~eV. Fitting hydrogenic series to their experimentally obtained peak energies resulted in estimated radii of 11.1 - 20~Å for the A$_1$ exciton \cite{Fortin_PhysicalReviewB_1975_11_2_905, Lee_OpticalandElectricalProperties_1976, Frindt_ProceedingsoftheRoyalSocietyofLondon.SeriesA.MathematicalandPhysicalSciences_1963_273_1352_69, Goto_JournalofPhysics-CondensedMatter_2000_12_30_6719, Khan_IlNuovoCimentoD_1983_2_3_665} and 5.3~Å for the B$_1$ exciton \cite{Lee_OpticalandElectricalProperties_1976}. Although those orbits are not particularly large, they are still greater than the lattice constants of 3.16~Å \cite{Wilson_AdvancesinPhysics_1969_18_73_193} in the$ xy$-plane of the crystals. Therefore, they may be treated approximately like Wannier-Mott excitons. The A$_1$ and B$_1$ electron-hole pairs have been attributed to band splitting due to interlayer interaction and spin-orbit coupling. While it was initially suspected that they originate from the $\Gamma$ \cite{Wilson_AdvancesinPhysics_1969_18_73_193, Bromley_PhysicsLettersA_1970_33_4_242, Bromley_JournalofPhysicsC-SolidStatePhysics_1972_5_7_759, Edmondson_SolidStateCommunications_1972_10_11_1085, Beal_JournalofPhysicsC-SolidStatePhysics_1972_5_24_3540, Tanaka_JournalofthePhysicalSocietyofJapan_1978_45_6_1899, Saiki_physicastatussolidi(b)_1978_88_2_607,Meinhold_physicastatussolidi(b)_1976_73_1_105, Boeker_PhysicalReviewB_2001_64_23_235305} or the $A$ point \cite{Kasowski_PhysicalReviewLetters_1973_30_23_1175}, recent investigations locate them at the K point. \cite{Coehoorn_PhysicalReviewB_1987_35_12_6203, Mak_PhysicalReviewLetters_2010_105_13_136805, Cheiwchanchamnangij_PhysicalReviewB_2012_85_20_205302, Molina-Sanchez_PRB_2013_88_4_45412, Mattheiss_PhysicalReviewB_1973_8_8_3719} A microscopic understanding of photoabsorption and photoluminescence  properties of MoS$_2$, which promise future applications,\cite{Xu_Chemicalreviews_2013_113_5_3766,Ganatra_ACSnano_2014_8_5_4074} requires a detailed knowledge of the exciton properties that has not been achieved, yet.

%EELS
There have been a number of transmission electron energy-loss spectroscopy (EELS) investigations of $2H$-MoS$_2$ in the past. \cite{Liang_PhilosophicalMagazine_1969_19_161_1031, Zeppenfeld_OpticsCommunications_1970_1_8_377,Bell_AdvancesinPhysics_1976_25_1_53,Disko_Ultramicroscopy_1987_23_3_313} However, based on our knowledge, none of them were capable of resolving the relatively narrow excitonic transitions. We are now able to present measurements that clearly show the A$_1$ and B$_1$ electron-hole pairs. In addition, EELS allows studying not only the electronic excitations as functions of energy but also as a function of momentum.  This can provide further information on the electronic properties of the materials under investigation. \cite{Knupfer_Physicalreviewletters_1999_83_7_1443, Schuster_Physicalreviewletters_2007_98_3_37402, Kramberger_Physicalreviewletters_2008_100_19_196803, Marinopoulos_Physicalreviewletters_2002_89_7_76402} Our measurements reveal such momentum dispersions of the two excitons, which permitted the calculation of their effective masses. Moreover, we will compare our results to the dispersion for monolayer material calculated in a recent theoretical study \cite{Wu_Phys.Rev.B_2015_91__75310}.

%Estimates of the width of direct gap ranges from 1.62 eV to 2.16 eV  

\section{EXPERIMENT}
\begin{figure} [h]
	\includegraphics {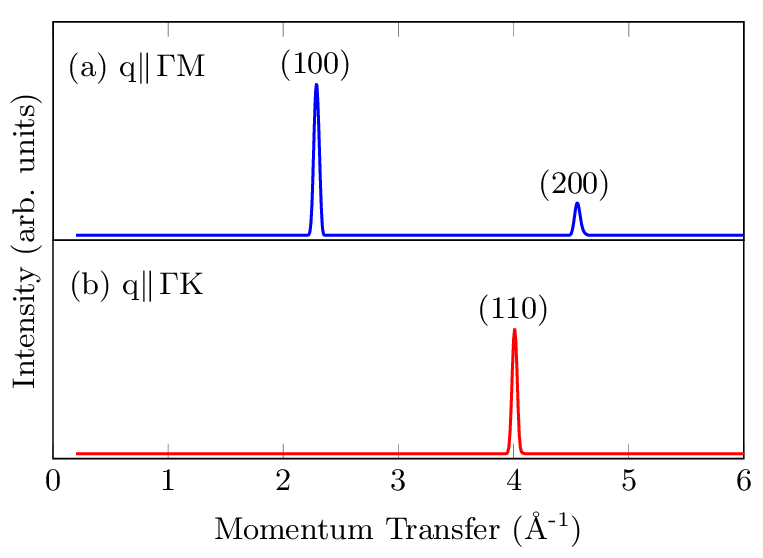}
	\caption{(Color online) Electron diffraction profiles of MoS$_2$ in (a) $\Gamma$M direction and (b) $\Gamma$K direction.}
	\label{fig1}
\end{figure}

Natural single crystal molybdenite was purchased from Manchester Nanomaterials. The bulk crystal was exfoliated $ex situ$ by repeated cleaving with adhesive tape until 100 nm thick films were produced. This choice of film thickness was a compromise balancing the negative effects of multiple scattering which increase with the number of layers with the count rate which becomes too low for practical purposes for thinner films. The thickness was estimated with the help of an optical microscope by visually comparing the transparency of the cleaved films to that of samples cut with a calibrated microtome. The film was placed on a platinum transmission electron microscopy grid. The measurements were done in a 172 keV transmission electron energy-loss spectrometer equipped with a helium flow cryostat (see Ref.~\citenum{Fink_AEEP_1989_75__121} and \citenum{  Roth_JournalofElectronSpectroscopyandRelatedPhenomena_2014_195__85} for a description of the spectrometer).  The energy and momentum resolution of the instrument were set to $\Delta E$~=~82 meV and $\Delta q$~=~0.04 Å$\textsuperscript{-1}$, respectively, and the sample temperature to 20 K. The film was aligned relative to the impinging electron beam by setting the spectrometer to momentum transfer values corresponding to the (100) and (110) Bragg peaks, respectively, and iteratively adjusting the azimuthal angle, the polar angle and to a very limited extent  the momentum transfer value until the detector signal in the respective direction was maximized. The resulting  spectra are presented in Fig.~\ref{fig1}. The locations of the (100) and (110) peaks are at approximately 2.29~Å$\textsuperscript{-1}$ and 4.00~Å$\textsuperscript{-1}$, respectively. The distinct peaks indicate a high degree of homogeneity of the crystal structure. 

% Figure 1:

The energy-loss spectra in the $\Gamma$M and $\Gamma$K directions were measured for various momentum transfer values between 0.07~Å$\textsuperscript{-1}$ and 0.3~Å$\textsuperscript{-1}$. The spectra were affected by the quasi-elastic line, reflecting the effects of elastic scattering. The line, which is approximately shaped like a Gaussian curve centered at zero, falls off quickly influencing only very low energy regions noticeably. Because of the relatively large separation between the elastic line peak and the first exciton, the effects of elastic scattering were removed from the measured data by shifting the intensity values of the spectra down by an amount representing the intensity at the lowest point of the fitted curve between the elastic line peak and the first exciton. For energies lower than the latter point, the intensities were set to zero. Moreover, the spectra were adjusted for the effects of multiple scattering.\cite{Livins_PRB_1988_38_8_5511}

\section{RESULTS AND DISCUSSION}
\begin{figure}
	\includegraphics {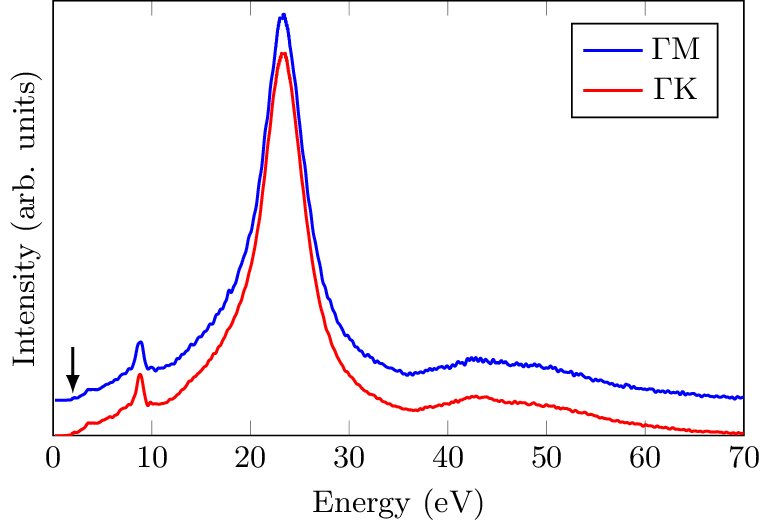}
	\caption{(Color online) Electron energy-loss spectra measured along the $\Gamma$M and $\Gamma$K directions with momentum transfer $q$ = 0.1~Å$\textsuperscript{-1}$ after removing the elastic line and multiple scattering effects. The spectra are offset along the intensity axis for clarity. The arrow indicates the position of the excitons.}
	\label{fig2}
\end{figure}

 Figure \ref{fig2} presents the energy-loss functions for a large energy range and a momentum transfer of $q$~=~0.1~Å$\textsuperscript{-1}$. The spectra imply a strong isotropy of the material in the $\Gamma$M and $\Gamma$K directions, which due to the hexagonal crystal symmetry is expected for low momentum transfers representing the optical limit \cite{Fink_AEEP_1989_75__121}. The main peak at 23.3~eV represents the volume plasmon, that is, the collective excitation of all 18 valence electrons per MoS$_2$ unit. The lower peak at 8.8~eV is thought to arises only from the oscillations of the electrons not involved in the ionic-covalent bonds between molybdenum and sulfur.\cite{Liang_PhilosophicalMagazine_1969_19_161_1031}  The energies of those two plasmons agree well with the 23.0 -~23.4~eV and 8.7 -~8.9 eV measured in previous EELS experiments.\cite{Liang_PhilosophicalMagazine_1969_19_161_1031,Zeppenfeld_OpticsCommunications_1970_1_8_377,Bell_AdvancesinPhysics_1976_25_1_53,Disko_Ultramicroscopy_1987_23_3_313}
The excitations of the 4p$_{1/2}$ and 4p$_{3/2}$ states of molybdenum are visible between 40 and 50~eV.

% Figure 2:

\begin{figure*}
	\includegraphics {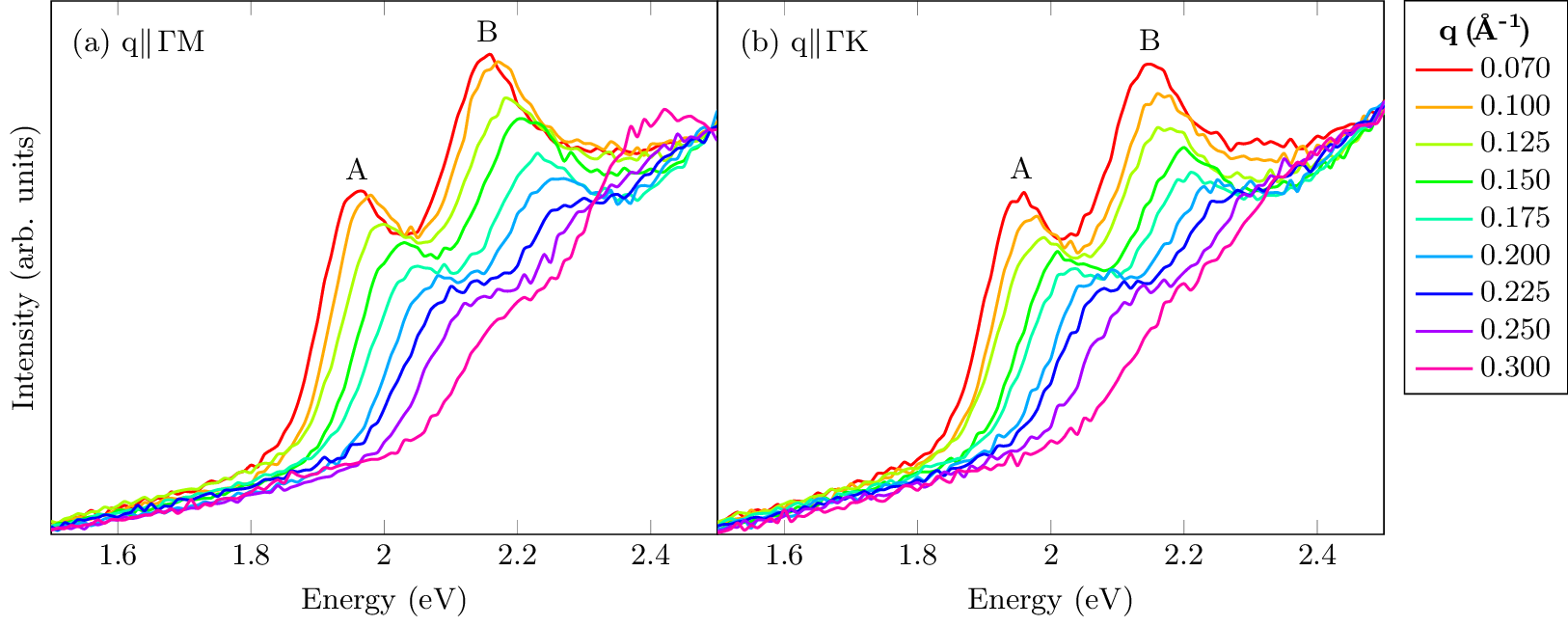}
	\caption{(Color online) Electron energy-loss spectra measured along the (a) $\Gamma$M and (b) $\Gamma$K direction with various momentum transfers showing the dispersion of excitons. The A$_1$ and B$_1$ exciton peaks for \textit{q} = 0.07~Å$\textsuperscript{-1}$  are labeled.}
	\label{fig3}
\end{figure*}

Close-up looks at the energy-loss spectra between 1.5 and 2.5~eV for the $\Gamma$M and $\Gamma$K directions are shown in Fig.~\ref{fig3}. The plots reflect the locations and dispersions of the A$_1$ and B$_1$ exciton peaks for momentum transfers ranging from 0.07~ Å$\textsuperscript{-1}$ through 0.3~Å$\textsuperscript{-1}$. For both directions, the exciton peaks exhibit a positive dispersion because their positions shift to higher energies as the momentum transfer increases. 
%Moreover, the energy difference between the A$_1$ and B$_1$ exciton peaks shrinks as the momentum transfer rises. For example, the energy split changed from 0.20~eV to~0.17 eV when the momentum transfer increased from 0.07~Å$\textsuperscript{-1}$ to 0.2~Å$\textsuperscript{-1}$ as demonstrated in Fig.~\ref{fig4}.

% Figure 3:

\begin{figure}
	\includegraphics {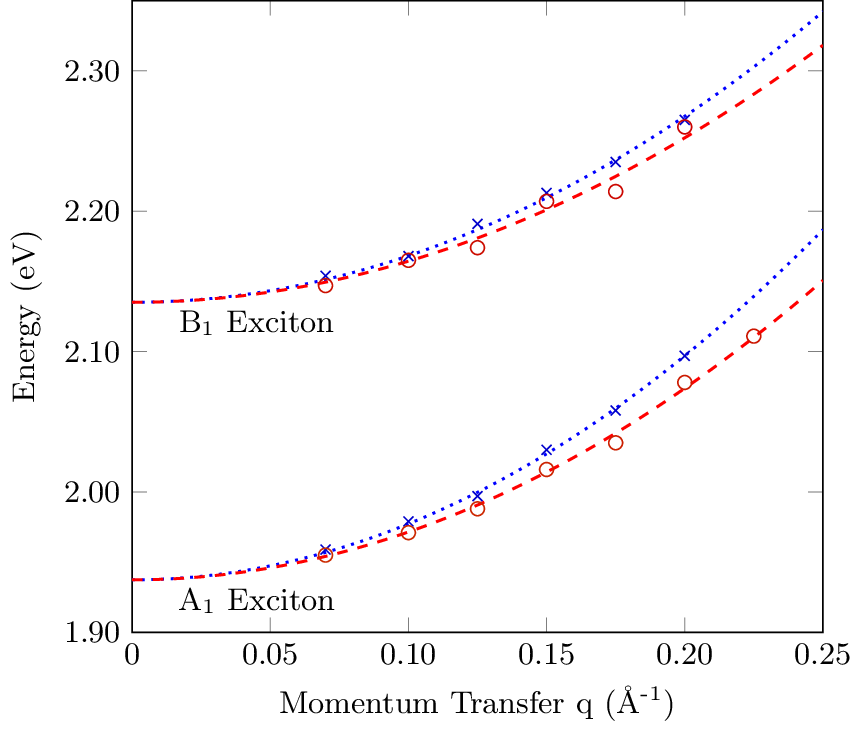}
	\caption{(Color online) Experimental dispersion of the energy-loss peak positions of the A$_1$ and B$_1$ excitons as function of momentum transfer in the $\Gamma$M (\textcolor{blue}{$\times$}) and $\Gamma$K (\textcolor{red}{$\medcircle$}) directions and fitted dispersion in the  $\Gamma$M (\textcolor{blue}{\hdashrule[0.5ex]{0.8cm}{0.7pt}{0.4mm}}) and $\Gamma$K (\textcolor{red}{\hdashrule[0.5ex]{0.8cm}{0.7pt}{0.9mm}}) directions. The peak positions were obtained by applying polynomial fits to the data presented in Fig. \ref{fig4} and validated quantitatively.}
%	\caption{Measured and fitted dispersion of the energy-loss peak positions of the A$_1$ and B$_1$ excitons as function of momentum transfer.}
	\label{fig4}
\end{figure}

Figure~\ref{fig4} exhibits the dispersion of the energy-loss peak positions related to the A$_1$ and B$_1$ excitons as function of momentum transfer. This information was extracted from the data presented in Fig.~\ref{fig3}. Given the approximate quadratic shape of the dispersion close to $q$~=~0, the effective mass approximation \cite{Wang_Ultramicroscopy_1995_59_1_109} (EMA) may be used to derive the effective mass of the excitons $m^*$ from the related energies $E(q)$ and momentum values $q$:

\begin{equation}
	E(q)=E(0) + \frac{\hbar^2}{2m^*}q^2.
	\label{equ:EMA}
\end{equation}

Equation~\ref{equ:EMA} was fitted to the experimental dispersion data and the common intersections of the $\Gamma$M and $\Gamma$K dispersion curves for the respective exciton peak at $q$~=~0. The latter constraint arises from the fact that all dispersion curves for a particular exciton should converge at the $\Gamma$ point of the Brillouin zone where the momentum transfer equals zero. The energies $E(0)$ at this point represent the transition energies of the excitons in the optical limit. The numerical results and the resulting effective exciton masses perpendicular to the c axis are listed in Table~\ref{tab:Disp}. It should be pointed out that the quality of the fits for the A$_1$ exciton dispersions is good while those for the B$_1$ exciton show somewhat larger deviations between the fitted dispersion functions and the measured energy-loss peak positions (see Fig.~\ref{fig4}). This is attributed to the fact that the B$_1$ exciton peaks are broader which makes the identification of the peak maximum location less precise. 

% Figure 4: Dispersion Fit

%Table: Calculated dispersion parameters and effective masses
\begin{table}
	\caption{Dispersion curve fitting results and effective exciton masses}
	\renewcommand{\tabcolsep}{0.2cm}
	\begin{tabular} {ccccc}
		\toprule
		\toprule
						& Momentum		&		& \multirow{2}{*}{ $\frac{\hbar^2}{2m^*}$}	& Effective\\
     		Exciton			& transfer		& $E(q=0)$	& 							& mass \\
      		peak				& direction		& (eV) 	& (eV\,Å$^2$)					& (m$_0$)\\
		\hline 

		\multirow{2}{*}{A$_1$}	&  $\Gamma$M	& 1.94	& 3.99						& 0.96\\
      						& $\Gamma$K	& 1.94	& 3.41 						& 1.12\\
      						&			&		& 							& \\
		\multirow{2}{*}{B$_1$}	& $\Gamma$M	& 2.14	& 3.31						& 1.15\\
      						& $\Gamma$K	& 2.14	& 2.93						& 1.30\\
		\hline 
		\bottomrule
	\end{tabular}
	\label{tab:Disp}
\end{table}

\subsection{Ground state energies}
The ground state energies $E(0)$ derived from the experimental data are 1.94 and 2.14~eV for the A$_1$ and B$_1$ excitons, respectively.\footnote{It should be pointed out that in contrast to optical spectroscopy, the zero momentum energy values $E(0)$ cannot be measured directly in an electron energy-loss spectroscope because all measurements have to be made with non-zero momenta. $E(0)$ was determined indirectly by fitting the EMA function to the exciton peak dispersion values and extrapolating this function to zero momentum as shown in Fig. 4.} In the temperature range from 4~K through 70~K,  peak energies of 1.91 -~1.9356~eV and 2.11 - 2.137 eV have been observed in other, mainly optical experiments for the A$_1$ and B$_1$ excitons, respectively.{\cite{Frindt_ProceedingsoftheRoyalSocietyofLondon.SeriesA.MathematicalandPhysicalSciences_1963_273_1352_69, Fortin_PhysicalReviewB_1975_11_2_905, Fortin_PhysicalReviewB_1975_11_2_905,Beal_JournalofPhysicsC-SolidStatePhysics_1972_5_24_3540, Frey_PRB_1998_57_11_6666, Evans_physicastatussolidi(b)_1968_25_1_417} Our findings are slightly above the values found in those previous investigations. There are a number of factors that may account for the differences. For example, EELS and optical measurements have different response functions.\footnote{The EELS response function is -Im$\frac{1}{\epsilon}$ with $\epsilon$ being the dielectric function.  \cite{Liang_PhilosophicalMagazine_1969_19_161_1031} In contrast, the response function of transmission experiments is $\omega \cdot \kappa$ with $\omega$ and $\kappa$ being the frequency and the imaginary part of the complex index of refraction, respectively.} As a consequence,  electron energy-loss spectra reflect excitations at somewhat higher energy values. We performed a Kramers-Kronig analysis and found that the calculated absorption spectra show the two excitons at energies that are about 0.02 eV lower than the EELS spectra. This places the exciton position well within the range found by optical investigations. In addition, differences may be due to different experimental conditions such as temperature and sample thickness. Exciton peaks tend to be blue-shifted as the temperature \cite{Frey_PRB_1998_57_11_6666, Fortin_PhysicalReviewB_1975_11_2_905, Khan_IlNuovoCimentoD_1983_2_3_665} or the sample thickness is decreased\cite{Li_arXivpreprintarXiv-1407.6997____}.

\subsection{Effective masses}
The dispersion curves and, therefore, the effective masses show a slight anisotropy. The dispersions are somewhat lower in the $\Gamma$K compared to the $\Gamma$M direction while the opposite is true for the effective masses (see Table \ref{tab:Disp}). The anisotropy is less than 15\%. Moreover, the A$_1$ exciton has a higher dispersion than the B$_1$ exciton as the values for $\hbar^2/({2m^*})$  in Tab. \ref{tab:Disp} indicate. That means that the energy difference between the peaks of the two exciton types shrinks as the momentum transfer rises. This is exemplified in Fig.~\ref{fig5}, where the energy split changes from 0.20~eV to 0.17~eV when the momentum transfer increases from 0.07~Å$\textsuperscript{-1}$ to 0.2~Å$\textsuperscript{-1}$. As a consequence, the mass of the A$_1$ exciton appears to be lower than that of the B$_1$ exciton. This general relationship, although more pronounced, is also observed in experimentally derived reduced exciton masses confirming our result. Based on absorption measurements, Evans calculated reduced masses of 0.31~m$_0$ for A$_1$ and 0.99~m$_0$ for B$_1$ in the xy-plane of the crystals.\cite{Lee_OpticalandElectricalProperties_1976} 
It should be stressed, however, that this comparison between effective and reduced masses is only a reasonability check. The numerical values are not directly comparable as they represent different concepts. While the effective mass is derived from the EMA in Eq. \ref{equ:EMA}, the reduced mass $\mu_\perp$ of an exciton perpendicular to the c axis is $\frac{1}{\mu\perp}=\frac{1}{m_{e\perp}^*}+\frac{1}{m_{h\perp}^*}$ were $m_{e\perp}^*$ and $m_{h\perp}^*$ are the electron and hole masses in the xy-plane, respectively.\cite{Evans_physicastatussolidi(b)_1968_25_1_417}  Experimentally, it is deduced from the spectral peak energies of an excitonic series similar to a hydrogenic series: $E_n=-\frac{\mu_\perp e^4}{\epsilon_0^2 2\hbar^2 n^2}$ where $E_n$ is the energy of the $n$th peak, e is the electron charge, $\epsilon_0$ is the permittivity of free space and $\hbar$ is the Planck constant.\cite{Lee_OpticalandElectricalProperties_1976}

%\footnote{The effective mass is derived from the EMA in Equ. \ref{equ:EMA}. It represents a parameter that if substituted for the free particle mass, makes the mathematical formulations of the free particle laws applicable to particles subject to forces. On the other hand, the reduced mass is a quantity that permitts considering the movement of the electron-hole system as the motion of a single particle with that mass. The reduced mass of an exciton perpendicular to the c axis $\mu_\perp$ is $\frac{1}{\mu\perp}=\frac{1}{m_{e\perp}^*}+\frac{1}{m_{h\perp}^*}$ where $m_{e\perp}^*$ and $m_{h\perp}^*$ are the electron and hole masses perpendicular to the c axis, respectively. \cite{Evans_physicastatussolidi(b)_1968_25_1_417} Experimentally, it is deduced from the spectral peak energies of an excitonic series similar to a hydrogenic series: $E_n=-\frac{\mu_\perp e^4}{\epsilon_0^2 2\hbar^2 n^2}$ where $E_n$ is the energy of the $n$th peak, e is the electron charge, $\epsilon_0$ is the permittivity of free space and $\hbar$ is the Planck constant.\cite{Lee_OpticalandElectricalProperties_1976}}

% Figure 5:
\begin{figure}
	\includegraphics {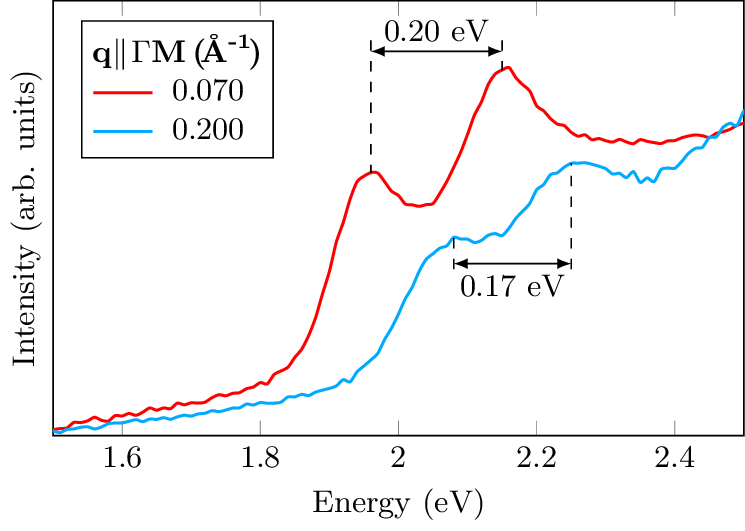}
	\caption{(Color online) Electron energy-loss spectra measured along the $\Gamma$M direction showing the reduction of the energy difference between exciton peak A$_1$ and B$_1$ with increasing momentum transfer.}
	\label{fig5}
\end{figure}

In a first approximation, the sum of the effective masses of an exciton's electron $m_e^*$ and hole $m_h^*$ may be used to validate the found exciton masses:

\begin{equation}
	m^*=m_e^*+m_h^*.
	\label{equ:FirstApprox}
\end{equation}

Based on theoretical band structure calculations, Yun et al. \cite{Yun_PhysicalReviewB_2012_85_3_33305} computed the effective electron and hole masses to be $m_e^*$~=~0.821~m$_0$ and  $m_h^*$~=~0.625~m$_0$ at the K point where the excitons originate. Plugging those values into Eq.~\ref{equ:FirstApprox} results in an effective exciton mass of 1.446~m$_0$ which is somewhat higher than our values in Table~\ref{tab:Disp}.  From their band structure model, Peelaers and Van de Walle \cite{Peelaers_PRB_2012_86_24_241401} calculated effective electron masses at the K point of 0.43~m$_0$ in the $\Gamma$ direction and 0.47~m$_0$ in the M direction. They found effective hole masses of 0.47 m$_0$ in the $\Gamma$ direction and 0.45~m$_0$ in the M direction at the same point. Averaging in respect to the directions produces effective masses of $m_e^*$~=~0.45~m$_0$ and $m_h^*$~=~0.46~m$_0$. Using Eq.~\ref{equ:FirstApprox}, the effective exciton mass would be approximately 0.91~m$_0$ which is lower than our estimates.

To account for the fact that the properties of a particular electron-hole pair might be somewhere between those of a Frenkel exciton and those of a Wannier-Mott exciton, Mattis and Gallinar\cite{Mattis_Prl_1984_53_14_1391} suggested a refinement of the effective mass approximation as presented in Eq.\ref{equ:FirstApprox}. They introduced a correction factor based on the kinetic energy $K_n$ in the $n$th bound exciton state and one-half the sum of the electron and hole bandwidths W resulting in a mass of an exciton in its $n$th bound state of:

\begin{equation}
	m_n^*=\frac{m_e^*+m_h^*}{1-\frac{K_n}{W}}.
	\label{equ:SecondApprox}
\end{equation}
 
Such a correction might be appropriate for MoS$_2$ because the excitons do not meet the perfect Wannier-Mott model due to their relatively small radii as indicated above. The exact kinetic energies $K_n$ of the observed excitons are not known. To obtain at least a rough estimate of the correction factor, we will assume that the hydrogenic virial theorem can be applied so that  the magnitude of the kinetic energy is comparable to the exciton binding energies.  Reported ground state binding energies for the A$_1$ and B$_1$ excitons range from 25 -~60~meV \cite{Yoffe_AnnualReviewofMaterialsScience_1973_3_1_147, Goto_JournalofPhysics-CondensedMatter_2000_12_30_6719, Beal_JournalofPhysicsC-SolidStatePhysics_1972_5_24_3540, Bordas_physicastatussolidi(b)_1973_60_2_505, Evans_ProceedingsoftheRoyalSocietyofLondon.SeriesA.Mathematicalandphysicalsciences_1965_284_1398_402, Cheiwchanchamnangij_PhysicalReviewB_2012_85_20_205302} and 130 -~136~meV, \cite{Lee_OpticalandElectricalProperties_1976, Evans_ProceedingsoftheRoyalSocietyofLondon.SeriesA.Mathematicalandphysicalsciences_1965_284_1398_402, Evans_ProceedingsoftheRoyalSocietyofLondon.SeriesA.MathematicalandPhysicalSciences_1967_298_1452_74, Evans_physicastatussolidi(b)_1968_25_1_417} respectively. Their mean values provide estimated kinetic energies of $K_1^A$~=~43~meV and $K_1^B$~=~134 meV  (superscripts A and B refer to the particular exciton).
To determine $W$ in Eq.~\ref{equ:SecondApprox}, the bandwidths were obtained from a number of theoretical band structure models (extracted from the applicable band structure plots in Ref. \citenum{Peelaers_PRB_2012_86_24_241401, Molina-Sanchez_PRB_2013_88_4_45412, Cheiwchanchamnangij_PhysicalReviewB_2012_85_20_205302}). The average half joint bandwidth was about $W$=1.6 eV. Consequently, the effective mass correction factors are 0.97 for the A$_1$ exciton and 0.92 for the B$_1$ exciton.
Applying them to the EMA estimate of 1.446~m$_0$ based on the values of Yun et al. \cite{Yun_PhysicalReviewB_2012_85_3_33305}  results in adjusted effective masses of $m_1^{*A}$~=~1.49~m$_0$ and $m_1^{*B}~=~$ 1.58~m$_0$ for the A$_1$ and B$_1$ excitons, respectively.  Making the same correction to $m^*$~=~0.91~m$_0$ leads to $m_1^{*A}$~=~0.94~m$_0$ and $m_1^{*B}$~=~0.99~m$_0$ which is still slightly lower than our computed values. Even after the adjustments, the effective exciton masses found in this work differ by up to 35\% from the numbers according to the EMA. We believe that our findings will allow the refinement of present band structure models to achieve a closer agreement of experimental and theoretical exciton effective mass values.

Wu et. al. \cite{Wu_Phys.Rev.B_2015_91__75310} calculated the dispersions of the A$_1$ and  B$_1$ excitons in monolayer $2H$-MoS$_2$. They found a splitting of both excitons into a mode with quadratic dispersion and a mode with linear dispersion. We fitted Eq. \ref{equ:EMA} to their dispersion values of the quadratic mode. \footnote{The dispersion data points were extracted from Fig. 2 of  Ref. \citenum{Wu_Phys.Rev.B_2015_91__75310}} The good fit showed that the general quadratic energy-momentum relationships agree quantitatively to our findings for the bulk material. In addition, the theoretical data reflect the disipation of the excitons at higher momentum values in agreement with our observations as reflected in Fig.~\ref{fig3}. The dispersion coefficients $\frac{\hbar^2}{2m^*}$ derived from the fit of Eq. \ref{equ:EMA} to the data of Wu et. al. are approximately 2.7 eV\,Å$^2$ and 2.5 eV\,Å$^2$ in the $\Gamma$M direction resulting in effective masses of 1.4~m$_0$ and 1.5~m$_0$ for the  A$_1$ and  B$_1$ excitons, respectively. Those masses are significantly higher than our experimental values for bulk $2H$-MoS$_2$ (see Table \ref{tab:Disp}). This is unexpected because previous theoretical studies found masses in monolayer material to be lower than in bulk. For example, approximate exciton masses based on Eq.~\ref{equ:EMA} and electron and hole masses published by Peelaers and Van de Walle \cite{Peelaers_PRB_2012_86_24_241401} are  0.84~m$_0$ for monolayers and 0.91~m$_0$ for bulk material. Using  Yun et al. \cite{Yun_PhysicalReviewB_2012_85_3_33305} electron and hole masses lead to exciton masses of 1.12~m$_0$ in monolayers and 1.45~m$_0$ in bulk. We believe that our findings can be used as an additional benchmark in the refinement of present band structure and exciton models which may lead to the reconciliation of the presented differences in effective exciton masses between bulk and monolayer material as well as experimental and theoretical values.

Since MoS$_2$ is an indirect semiconductor, we also looked for signatures of indirect excitons but were unable to detect them.

\section{summary}
The transmission electron energy-loss spectra of bulk $2H$-MoS$_2$ were measured. The A$_1$ and B$_1$ excitons peaks and their momentum transfer dispersion in the $\Gamma$M and $\Gamma$K directions were observed. The dispersions were found to be positive and approximately quadratic. They showed a slight anisotropie. Moreover, the exciton masses were calculated and in reasonable agreement with expectations. 

\begin{acknowledgments}
We thank R. H\"ubel, S. Leger and M. Naumann for their technical assistance.
\end{acknowledgments}

\bibliography{Dichalcogenide}

\end{document}